\documentclass[doublecol,figures]{epl2}
\usepackage{graphicx}
\usepackage{subfig}
\usepackage{array,amsmath}
\usepackage{appendix}
\usepackage{comment}
\usepackage{amssymb}
\usepackage{cuted}

\title{Optimal detuning and gate time for strongly driven single-qubit gates beyond the rotating-wave approximation}
\shorttitle{Frequency and time corrections for single-qubit gates beyond the rotating-wave approximation}
\date{\today}
\author{Alejandro D. Bendersky\inst{1,2} \and Sergio S. Gomez \inst{1,2} \and Rodolfo H. Romero \inst{1,2}\thanks{Corresponding author: \email{rhromero@exa.unne.edu.ar}}}

\shortauthor{A. D. Bendersky et al}

\institute{                    
  \inst{1} Instituto de Modelado e Innovación Tecnológica, CONICET, Avenida Libertad 5500, 3400 Corrientes, Argentina.\\
  \inst{2} Facultad de Ciencias Exactas y Naturales y Agrimensura, Universidad Nacional del Nordeste, 
  Avenida Libertad 5500, 3400 Corrientes, Argentina
  }

\abstract{
Single-qubit quantum gates are implemented as the time-controlled unitary evolution of a two-level system under harmonic driving. 
The rotating-wave approximation (RWA) provides closed-form gate parameters for weak coupling. Larger amplitudes, desirable to speed up the dynamics, make the RWA-designed gates inaccurate because of the neglected counter-rotating term, which also shifts the resonance (Bloch-Siegert shift). 
Here we study theoretically the infidelity of gates beyond the RWA, using the effective Magnus-Taylor Hamiltonian through second order in the inverse of the driving frequency. 
By detuning the driving frequency by an amplitude-dependent shift of Bloch-Siegert type and rescaling the operation time, the RWA gate conditions are recovered with corrected parameters. 
Numerical solutions of the Schr\"odinger equation show that these corrections lower the gate infidelity by several orders of magnitude, extending the range of usable amplitudes.
}
\begin{document}
\maketitle

\section{Introduction}
Quantum gates --the unitary operators acting on individual or assemblies of two-level quantum systems-- are usually implemented as the unitary evolution, during a definite time, generated by an externally controlled Hamiltonian. 
Single-qubit gates are routinely implemented by driving a two-level system with a harmonic field of controlled amplitude, frequency and duration. 
For weak driving, the rotating-wave approximation (RWA) turns the problem into a static one in the rotating frame and yields closed-form gate times and amplitudes, which is why it underlies gate design in magnetic resonance, superconducting circuits, spin qubits and quantum dots \cite{Sacolick10, Zhang18, Bendersky26}. 
The price is the neglect of the counter-rotating component of the drive. 
As the driving amplitude grows, this term shifts the resonance \cite{Bloch40, Cohen73}, modifies the Rabi oscillations \cite{Zelo97, Nori07, Tuorila10, Lu12, Yan15} and introduces coherent errors in the gates. 
The Bloch-Siegert (BS) shift, i.e. the frequency shift of the spectral resonance, has been observed in a wide variety of systems \cite{Sie17, Pietikainen17, Li18}; deviations from the RWA have also been reported for a single strongly driven spin \cite{Fuchs09}, and the harmonically driven two-level problem has motivated both analytical and numerical approaches \cite{Shirley65, Grifoni98, Barnes12, Barnes13, Giscard20}.

The relevance of these effects is practical. 
Fast gates require strong driving to outpace decoherence, whereas fault-tolerant protocols require gate errors below roughly $10^{-3}$ \cite{Fowler12}. 
Since the RWA infidelity grows as a power law in the ratio between the driving amplitude and frequency, reaching this threshold within the RWA forces weak and therefore slow driving, which exposes the qubit to noise for longer \cite{Yang17, Yang19, Mickelsen23}. 
Strategies to deal with the BS shift and other non-RWA errors include its compensation or elimination through dedicated experimental schemes \cite{Pradhan09, Zhang18, Sudyka19}, pulse shaping and optimized control \cite{Yang17, Yang19, Motzoi09}, drive corrections that minimize the counter-rotating contribution \cite{Zwanenburg25}, and systematic high-frequency expansions, either Magnus-based \cite{Blanes09, DiVincenzo20}, of Floquet-Magnus type with controlled error \cite{Bukov15, Dey25}, or based on multitimescale averaging \cite{Barajas26}. 

In this work we derive a closed-form prescription for the corrected driving frequency and operation time of the complete $Z$, $X$ and Hadamard set within a unitarity-preserving effective-Hamiltonian framework, together with an analytical explanation of the power-law growth of the RWA gate error.

We use the dynamics of the harmonically driven two-level BS Hamiltonian
\begin{equation}
    H_{\rm BS}(t)=\frac{\omega_0}{2}\sigma_z+A\sin(\omega t) \sigma_x,
    \label{H_BS}
\end{equation}
 to define quantum computing gates; here $\omega_0$ is the energy gap between both levels, $A$ is the amplitude of the harmonic coupling oscillating with frequency $\omega$ and $\sigma_x$, $\sigma_z$ are the Pauli matrices. We take $\hbar=1$ throughout this work. 

Single-qubit gates $G$ are obtained from the time-dependent dynamics of a single two-level system. Given a Hamiltonian $H(t)$, its evolution operator satisfies $i\dot{U}= H(t) U$, where the dot represents the time derivative. 
After solving for $U(t,0)$ with $U(0,0)=I$, a convenient choice of the gate operation time $t=t_g$ and particular values for the parameters in $H(t)$, allows one to implement the various single-qubit gates $G$ by the condition that the unitary evolution reproduce the effect of the gate, i.e.,  $U(t_g,0)|\psi(0)\rangle=G|\psi (0)\rangle$. 
Then, by properly choosing the operation time $t_g$ and particular values of the ratio $A/\omega_0$, accurate implementations of various gates $G$ are obtained.

In this Letter we use the second-order Magnus-Taylor effective Hamiltonian \cite{DiVincenzo20} of the harmonically driven two-level system to define $Z$, $X$ and Hadamard gates that remain accurate for large driving amplitudes. 
Our main results are: (i) the RWA gate parameters are corrected analytically, through an amplitude-dependent shift of the driving frequency, of Bloch-Siegert type, and a rescaling of the operation time, eqs. (\ref{eq:gate-times-corrected}); (ii) the infidelity of RWA-designed gates is shown to follow a power law in $A/\omega$, with an analytical estimate of its origin, eqs. (\ref{power-law-error}); and (iii) numerical solutions of the Schr\"odinger equation with the corrected parameters show an improvement of the gate fidelity by several orders of magnitude, extending the usable range of amplitudes up to $A/\omega\lesssim 1$ (Fig. \ref{fig:time-frequency-shift}).

\section{Infidelity of RWA-derived gates}
As stated above, a satisfactory analytical solution to the $H_{\rm BS}(t)$ dynamics can be obtained for a small $A/\omega$ ratio. 
In this case, the widely used approach is the rotating wave approximation (RWA), that arises from noting that the coupling term consists in the superposition of two counter-rotating driving oscillations $\sim\exp{(\pm i \omega t)}$. Thus, by transforming $H_{\rm BS}(t)$ to a reference frame rotating with the same angular frequency $\omega$, one of them becomes static and frequency-independent, while the other turns into a time-dependent coupling with a higher frequency $2\omega$. The approximation consists in neglecting the high frequency counter-rotating term under the assumption that its contribution averages to zero in time. That assumption provides analytical solutions $U_{\rm RWA}(t,0)$ of the dynamics for weak coupling, i.e. for small amplitudes.

For instance, RWA provides an adequate description of the dynamics of spins in semiconductor quantum dots for small coupling amplitudes and can be used to model quantum gates for qubits consisting in spin-orbital states \cite{Bendersky26}. 
In that system, $A$ is the tunneling amplitude and the resonant driving  ($\omega=\omega_0$) during times $t_g=\tau_Z, \tau_X, \tau_H$ generates, respectively, $Z$, $X$ and Hadamard $H=(Z+X)/\sqrt{2}$ quantum gates for the particular values of $A/\omega$ given by ($n, k$ integer numbers)
%------------------------
\begin{subequations}
    \label{eq:gate-times}
    \begin{align}
      Z&=U_{\rm RWA}(\tau_Z), \hspace{0.3cm} \tau_Z = \frac{2n\pi}{A}, \hspace{1.4cm}  \frac{A}{\omega_0} =\frac{2n}{2k+1},  \\
      X&=U_{\rm RWA}(\tau_X), \hspace{0.3cm} \tau_X = \frac{(2n+1)\pi}{A}, \hspace{0.5cm}  \frac{A}{\omega_0} = \frac{2n+1}{2k+1}, \\
      H&=U_{\rm RWA}(\tau_H), \hspace{0.3cm} \tau_H = \frac{(4n+1)\pi}{2A}, \hspace{0.5cm}  \frac{A}{\omega_0} = \frac{4n+1}{2(2k+1)}.
    \end{align}
\end{subequations}

The fidelity $F(\theta,\varphi) = |\langle \psi|G^\dagger U|\psi\rangle|^2$ is a measure of the accuracy of the representation of $G$ by $U(t_g)$ through their action on an arbitrary state $|\psi(\theta,\varphi)\rangle$, and its average over the Bloch sphere ${\cal S}(\theta,\varphi)$ --the manifold of pure qubit states-- is 
\begin{equation}
    F_G = \frac{1}{6}\left[\textrm{Tr}(M_G M_G^\dagger)+|\textrm{Tr} M_G|^2\right] = \frac{1}{3}\left(1+\frac{1}{2}|\textrm{Tr} M_G|^2\right),
    \label{def-fidelity}
\end{equation}
where $M_G=G^\dagger U$ and, due to the unitarity of $G$ and $U$, $\textrm{Tr} (M_G M_G^\dagger)=2$. \\

\begin{figure}[h]
    \centering
  \begin{tabular}{cc}
    \subfloat{\includegraphics[width=0.45\linewidth]{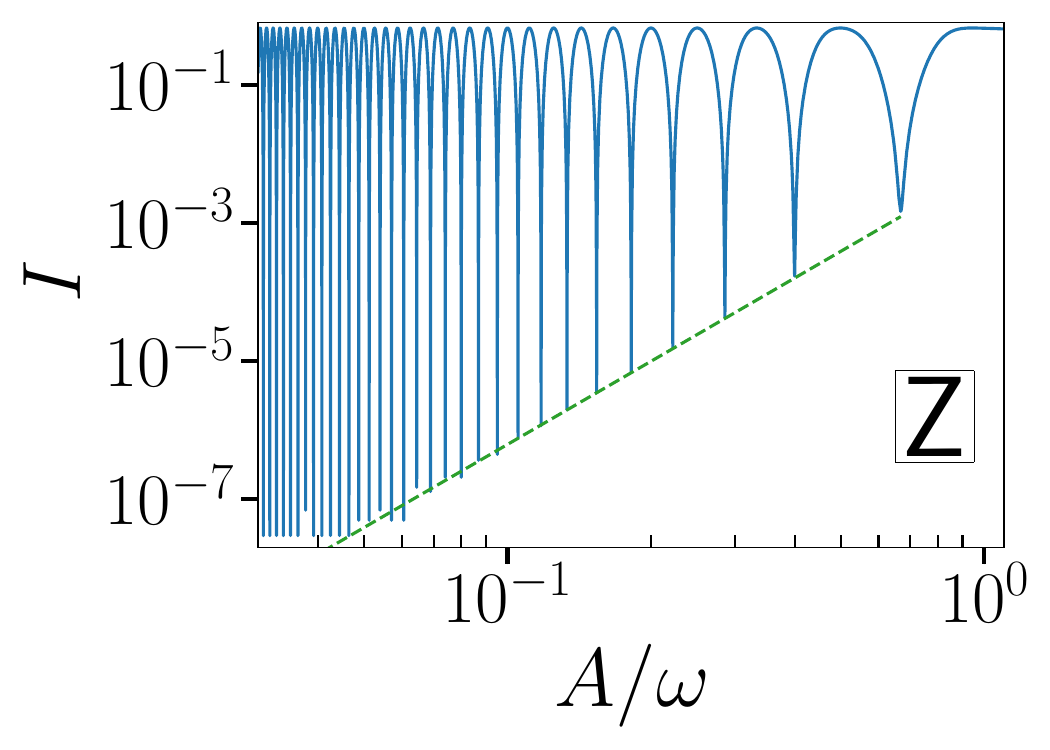}} & \hspace{-0.5cm}
    \subfloat{\includegraphics[width=0.45\linewidth]{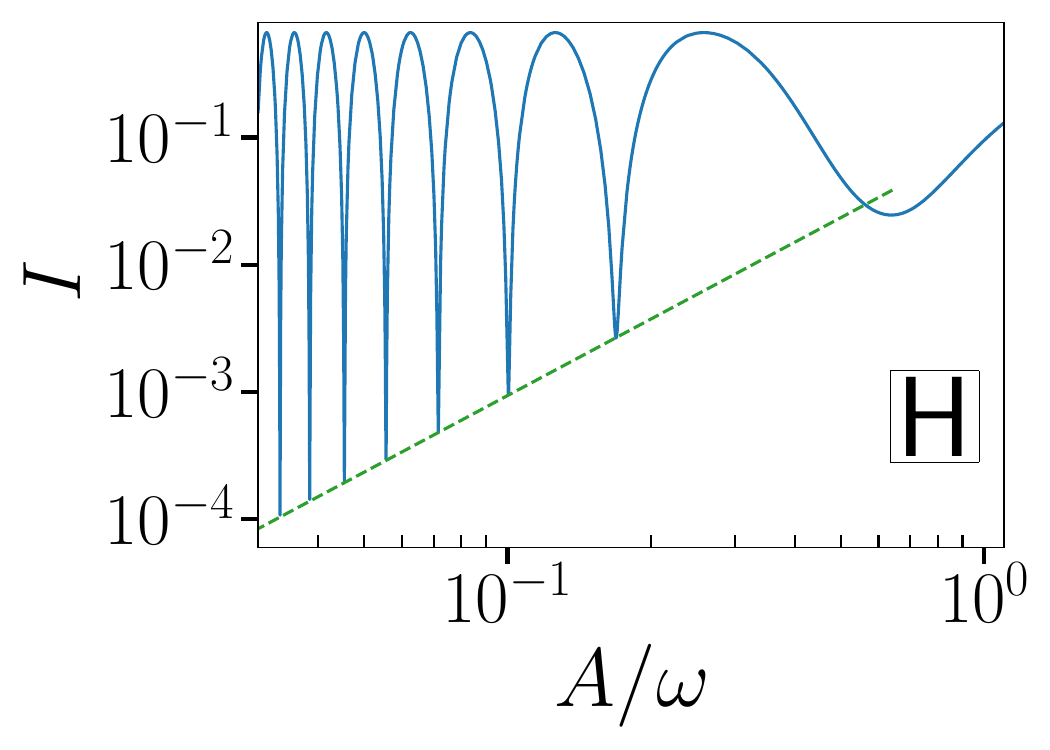} }
    \end{tabular}
    \caption{Infidelity of $Z$ and $H$ gates as a function of the relation $\lambda=A/\omega$ from the numerically calculated evolution operator for the Hamiltonian $H_{\rm BS}(t)$. The driving is taken in resonance with the energy gap ($\omega=\omega_0$) and the operation time is taken from the RWA prescription, $t_g=\tau_Z, \tau_H$ [eq. (\ref{eq:gate-times})]. The green dashed line are power-law fits at the minima of the dips: $\sim(A/\omega)^4$ for $Z$, and $\sim(A/\omega)^2$ for Hadamard gate.}
    \label{fig:Num-infidelity-lambda}
\end{figure}

Harmonic driving with large amplitudes decreases the accuracy of the representation of $G$ by $U_{\rm RWA}$. That is, the use of the RWA-derived parameters [eq. (\ref{eq:gate-times})] worsens the $U$-represented gate as $A/\omega$ increases and, therefore, the approximation becomes less valid.  Fig. \ref{fig:Num-infidelity-lambda} shows the infidelity $I=1-F$ calculated with eq. (\ref{def-fidelity}) for the $Z$ and Hadamard $H=(Z+X)/\sqrt{2}$ gates represented by the evolution operator $U$ obtained from numerically solving the Schr\"odinger equation for $H_{\rm BS}(t)$, evaluated at resonance ($\omega=\omega_0$) and evolving during $t_g=\tau_Z, \tau_H$, eq. (\ref{eq:gate-times}). Notice that the loss of accuracy, as measured by the minima of $I$ at the dips (green dashed line), increases with $A$ as $(A/\omega)^4$ and $(A/\omega)^2$ for $Z$ and $H$, respectively. Therefore, the $Z$ gate is less sensitive than $H$ to the increase of $A$, as long as $A<\omega$. Since gate errors must remain below $\approx10^{-3}$ for fault-tolerant operation \cite{Fowler12}, these power laws set an upper bound to the usable amplitude of RWA-designed gates and, therefore, a lower bound to their operation time. The origin of the different exponents is analyzed below, once the effective evolution operator is presented.
%------------------------
%=========================

Post-RWA approaches aim to provide a better estimation of the effect produced by the neglected counter-rotating term, and high-frequency expansions offer a systematic route to this goal \cite{Bukov15}. 
One of them is the Magnus-Taylor expansion \cite{DiVincenzo20}. It combines the Magnus expansion \cite{Blanes09} of the unknown evolution operator, represented in terms of a product of unitary factors to warrant its unitarity, with a Taylor expansion of the resulting time-dependent operators around a given time to get a time-independent unitary approximate evolution operator. 
We will define quantum gates in terms of evolution operator of that effective Hamiltonian.

\section{Magnus-Taylor-derived gates}
The Magnus-Taylor expansion is a perturbative method that ensures unitary evolution through any order in $1/\omega$, and allows larger amplitudes via a Taylor expansion in powers of $A$. We use here the second-order effective Hamiltonian $\widetilde{H}$ derived from that method in Ref. \cite{DiVincenzo20} and calculate its evolution operator to produce quantum gates. Briefly, $\widetilde{H}$ is obtained as follows. 

Starting from the Hamiltonian $H_{\rm BS}(t)$ [eq. (\ref{H_BS})] and transforming to a frame rotating with the driving frequency, i.e. $H_{\rm rot} = R^\dagger H R -i R^\dagger dR/dt$$= \exp(i\sigma_z\omega t/2)H\exp(-i\sigma_z \omega t/2)-(\omega/2)\sigma_z$, the Hamiltonian $H_{\rm rot}$ contains time-independent terms and driving terms with frequency $2\omega$. The effective Hamiltonian is defined as that  whose evolution operator produces the exact dynamics within an interval $\pi/\omega$, the drive period in the rotating frame; then, applying the Magnus-Taylor expansion up to second order in $1/\omega$ \cite{DiVincenzo20}, for our case of constant amplitude $A$, becomes an effective time-independent Hamiltonian $\widetilde{H}=H_0+H_1+H_2+{\cal O}(1/\omega^3)$, where $H_n\sim {\cal O}(1/\omega^n)$ \cite{note}.

$H_0$ gives the time evolution within RWA, while $H_1$ and $H_2$ are the linear and quadratic corrections in $A/\omega$ to the counter-rotating field, respectively. They also depend on the frequency detuning $\Delta=\omega_0-\omega$ between the two-level gap $\omega_0$ and that from the driving field; $0\leq\beta_0<2\pi$, is a gauge parameter depending on the initial time ($\beta_0=2\omega t_0$) for the fundamental Magnus interval $(t_0,t_0+\pi/\omega)$ that determines the `stroboscopic' coincidences with the exact evolution with period $\pi/\omega$ \cite{DiVincenzo20}.

We collect the various terms in $H_0$, $H_1$ and $H_2$ depending on each $\sigma_i$ component ($i=x, y, z$), to write the effective Hamiltonian as $\widetilde{H} = h_x\sigma_x + h_y\sigma_y +h_z\sigma_z $, where the $h_i$ are given in terms of the ratios $\lambda=A/\omega$ and $\delta=\Delta/\omega$ describing the relative importance of the amplitude $A$ and the detuning $\Delta$ with respect to the driving frequency $\omega$. The explicit general forms of $h_x$, $h_y$ and $h_z$ are given in the Appendix.\\

%%%%%%%%%%%%%%%%%%%%%%%%%%%%%%%%%%%%%
The unitary evolution of states driven by $\widetilde{H}$ is solved exactly by introducing the parametrization 
%-------------------------
\begin{subequations}
\begin{align}
    h_x &= \Omega\sin \theta\cos\varphi, \\
    h_y &= \Omega\sin \theta\sin\varphi, \\
    h_z &= \Omega\cos \theta, \\
    \Omega &= (h_x^2+h_y^2+h_z^2)^{1/2}, \label{Omega}
\end{align}
 \label{Omega-def}
\end{subequations}
%===================
such that $\widetilde{H}$ becomes
%=============================
\begin{equation}
   \widetilde{H} = \Omega  \left(
\begin{array}{cc}
  \cos \theta &  e^{-i\varphi}\sin \theta  \\
  e^{i\varphi}\sin \theta & -  \cos \theta \\
\end{array}
\right).
\label{H-parametric}
\end{equation}
%=============================
Reciprocally,
\begin{equation}
    \cos \theta = \frac{h_z}{\Omega},  \hspace{1cm}  \tan\varphi = \frac{h_y}{h_x}. \label{def-theta-varphi}
\end{equation}
%--------------------------------
It can be shown after some algebra, that while $h_i(\beta_0)$ depends on $\beta_0$ [eqs. (\ref{hx-general})--(\ref{hz-general})], $\Omega$ does not. Thus, the dynamics is gauge-independent or, equivalently, insensitive to the initial time. By choosing $\beta_0=0$, we get 
\begin{subequations}
\label{eqs:hx-hy-hz}
    \begin{align}
            h_x &=  0 , \label{hx approx} \\  
    h_y &=   - \frac{A}{2} \left( 1  - \frac{5}{16}\lambda^2 - \frac{\Delta}{2\omega} \right) \label{hy approx} \\
   h_z &=  \frac{\Delta}{2}\left( 1   -\frac{\lambda^2}{2} \right) + \frac{3}{8}A\lambda. \label{hz approx} 
    \end{align}
\end{subequations}
The RWA limit corresponds to the small-amplitude driving ($\lambda\rightarrow 0$), i.e. $\widetilde{H}_{\rm RWA}=-(A/2)\sigma_y+(\Delta/2)\sigma_z$. 
From this, we see the effective Magnus-Taylor Hamiltonian $\widetilde{H}$ is formally identical to $\widetilde{H}_{\rm RWA}$ if $h_y$ is considered as the lowest-order corrected amplitude due to the counter-rotating term neglected in RWA; similarly, $h_z$ would represent the effective detuning for the Magnus-Taylor Hamiltonian, so that even for resonant driving ($\Delta=0$) there is an effective shift $(3/8)A\lambda$ given by the second term in (\ref{hz approx}); the detuning $h_z$ can be suppressed through ${\cal O}(\lambda^2)$, by taking $\Delta= {\Delta_{\rm BS}}$, where  
%------------------------
\begin{equation}
    {\Delta_{\rm BS}}=-\frac{3}{4}A\lambda=-\frac{3}{4}\lambda^2\omega.
    \label{BS-shift}
\end{equation}
%------------------------
Thus, a frequency shift $\omega_0-\omega =\Delta_{\rm BS}$ in the true Hamiltonian $H_{\rm BS}$ corresponds to resonant driving $h_z=0$ in the effective one $\widetilde{H}$. Writing explicitly eqs. (\ref{eqs:hx-hy-hz}) in terms of $\Delta_{\rm BS}$, 

\begin{subequations}
\label{eqs:hy-hz-BS-shift}
\begin{align}
    h_y &= -\frac{A}{2}\left(1-\frac{5}{16}\lambda^2 +\frac{3}{8}\lambda^2\frac{\Delta}{\Delta_{\rm BS}}\right), \\
    h_z &= \frac{1}{2}(\Delta-\Delta_{\rm BS}),
    \end{align}
\end{subequations}

Notice also that $h_x=0$ and $h_y<0$ set $\varphi=-\pi/2$ in eq. (\ref{H-parametric}), leading to the parameterized expression  $\widetilde{H}=\Omega(\sigma_z\cos\theta-\sigma_y\sin\theta)$. Alternatively, it can be interpreted as the Hamiltonian of a spin 1/2 in a magnetic field $(h_x,h_y,h_z)=\Omega(0,-\sin\theta,\cos\theta)$, where $\theta$ represents the tilt of the field axis with respect to the $z$ axis, produced by the correction terms, see Fig. \ref{fig:time-frequency-shift}a.

Furthermore, from eq. (\ref{Omega}), replacing eqs. (\ref{eqs:hx-hy-hz}), expanding in $\lambda$ up to ${\cal O}(\lambda^2)$, and substituting eq. (\ref{BS-shift}), the energy eigenvalues of $\widetilde{H}$ are $\pm \Omega$   [Appendix, eq. (\ref{Omega-2nd-order})]
%======================== 
%___________________________
\begin{eqnarray}
    \Omega &\approx& \frac{A}{2}\left[1-\frac{\lambda^2}{32}+\frac{\Delta}{4\omega}+\frac{\Delta^2}{2A^2}\right] \nonumber \\
     &=& \frac{A}{2}\left(1-\frac{\lambda^2}{32}-\frac{3\lambda^2}{16}\frac{\Delta}{\Delta_{\rm BS}}+\frac{9\lambda^2}{32}\frac{\Delta^2}{\Delta_{\rm BS}^2}\right)  
    \label{Rabi-correction}
\end{eqnarray}
%----------------------------
Two remarks on the meaning of $\Delta_{\rm BS}$ are in order. First, $h_z$ depends on the gauge parameter $\beta_0$ [eq. (\ref{hz-general})] whereas $\Omega$ does not; the coefficient $3/4$ in eq. (\ref{BS-shift}) is therefore tied to the choice $\beta_0=0$, i.e., to a drive $\sin\omega t$ starting at $t=0$ and to gates evaluated at the stroboscopic times $t_g=(2m+1)\pi/\omega$ (see $\sin(\omega t_g/2)=\pm1$ below). $\Delta_{\rm BS}$ must thus be understood as the optimal gate detuning for this protocol. Second, it is not the shift of the spectral resonance usually associated with Bloch and Siegert \cite{Bloch40, Cohen73}, which is a property of the dressed spectrum and cannot depend on the initial phase of the drive. Extremizing $\Omega$ in eq. (\ref{Rabi-correction}) with respect to $\Delta$ gives $\Delta=-A\lambda/4$, the same value obtained when the coefficient $c_z$ of eq. (\ref{hz-general}) is replaced by its average over $\beta_0$, and equal to the standard shift $A^2/4\omega$ for the Rabi frequency $A$ implied by $\widetilde{H}_{\rm RWA}$.

Finally, the evolution operator $\widetilde{U}(t,0) = \exp(-i \widetilde{H}t)$ is transformed back to the laboratory frame, $U(t,0)=R(t) \widetilde{U}(t,0) R^\dagger(0)$, to get the propagator for the Magnus-Taylor approximation
 %%%%%%%%%%%%%%%%%%%%
%---------------------------------
\begin{strip}
\begin{equation}
    U(t,0) = \left(\begin{array}{cc}
 e^{-i\omega t/2}\left(\cos{\Omega t} -i\sin{\Omega t} \cos\theta\right) &
   -i  e^{-i (\varphi+\omega t/2) } \sin \Omega t \sin \theta  \\
-i e^{i (\varphi+\omega t/2) } \sin \Omega t \sin \theta & 
e^{i\omega t/2}\left(\cos{\Omega t} +i\sin{\Omega t} \cos\theta\right)  \\
    \end{array}
    \right).
    \label{U-lab}
\end{equation}
\end{strip}
%---------------------------------
The conditions for $U$ to represent the $Z$, $X$ and $H$ gates are derived from the expression of the fidelities. Using eq. (\ref{def-fidelity}), we get 
%------------------------------------------------
\begin{subequations}
\label{eq:fidelity} 
\begin{align}
    F_Z  &= \frac{1}{3}\left( 1+ 2 \left[\sin \left(\omega t/2\right) \cos \Omega t + \cos\theta \cos \left(\omega t/2 \right) \sin \Omega t\right]^2\right) \label{Fz} \\
    F_X  &=   \frac{1}{3}\left[ 1+2 \sin^2(\omega t/2) \sin^2 \theta \sin^2\Omega t \right], \label{Fx} \\
    F_H  &=  \frac{1}{3}\left[ 1+ \left(\sin \Omega t \left[\sin\theta \sin (\omega t/2)  \right. \right. \right. \nonumber \\
         & \hspace{0.2cm}  + \left. \left. \left. \cos\theta \cos (\omega t/2)\right]+\sin (\omega t/2) \cos \Omega t\right)^2\right]. \label{Fh}
\end{align}
\end{subequations}
%------------------------------------------
It can be seen that in the limit $\lambda\rightarrow 0$, $U(t)$ becomes $U_{\rm RWA}(t)$, which gives a perfect representation ($F=1$) of the gates at resonance ($\Delta=0$, $\Omega=A/2$) at the specific times $t$ and $A/\omega$ ratios given in eqs. (\ref{eq:gate-times}). Notice that the on-resonant condition $\Delta=0$ entails that the RWA fictitious magnetic field $\mathbf{\Omega}_R/2=-\sigma_y A/2$ (Fig. \ref{fig:time-frequency-shift}a) lies along the $-y$ axis, that is $\cos\theta=0$, from which the gate conditions (\ref{eq:gate-times}) are derived.\\

Now we show that perfect fidelity can also be achieved for the evolution operator $U(t)$ at finite $\lambda$. 
Analogously to the RWA case, it corresponds to the vector $\mathbf{\Omega}$ lying along $-y$ (Fig. \ref{fig:time-frequency-shift}a) with $\cos\theta=0$, i.e., $h_z=(\Delta-\Delta_{\rm BS})/2=0$.
Therefore, by taking a detuning $\Delta=\Delta_{\rm BS}$ in the effective $\widetilde{H}$, the energy eigenvalues become $\Omega= (A/2)(1+\lambda^2/16)$ [eq. (\ref{eqs:hy-hz-BS-shift})] and $U(t)$ takes the same form as the RWA one with the substitutions  $(A/2)\tau_g\rightarrow\Omega t_g$ and $\omega_0\rightarrow\omega=\omega_0-\Delta_{\rm BS}$. 

Accordingly, the conditions for $U$ to represent the gates through ${\cal O}(\lambda^2)$ can be directly written down from eqs. (\ref{eq:gate-times}):
%------------------------
\begin{subequations}
    \label{eq:gate-times-corrected}
    \begin{align}
      Z&=U(t_Z), \hspace{0.5cm}   \frac{2\Omega}{\omega_0-\Delta_{\rm BS}} =\frac{2n}{2k+1},  \\
      X&=U(t_X), \hspace{0.5cm}   \frac{2\Omega}{\omega_0-\Delta_{\rm BS}} = \frac{2n+1}{2k+1}, \\
      H&=U(t_H), \hspace{0.5cm}   \frac{2\Omega}{\omega_0-\Delta_{\rm BS}} = \frac{4n+1}{2(2k+1)}, \\
      t_g &= \frac{\tau_g}{1+\lambda^2/16}\approx \tau_g(1-\lambda^2/16), 
    \end{align}
\end{subequations}
with $g=Z, X, H$.

Equations (\ref{eq:gate-times-corrected}) hold through ${\cal O}(\lambda^2)$. With $\Delta=\Delta_{\rm BS}$ the effective detuning does not vanish exactly: from eq. (\ref{hz approx}), $h_z\approx(3/16)A\lambda^3$, and the terms of order $1/\omega^3$ neglected in $\widetilde{H}$ contribute at order $A\lambda^3$ as well. These residuals, together with the finite precision with which $\omega_0$ and $A$ are known in practice, limit the fidelity of the corrected gates at large $\lambda$. Notice also that the amplitudes $A/\omega_0$ allowed by eqs. (\ref{eq:gate-times-corrected}) form the same discrete set as the RWA conditions, eqs. (\ref{eq:gate-times}), with $A/\omega_0$ replaced by $2\Omega/(\omega_0-\Delta_{\rm BS})$: for given integers $(n,k)$ this means $\lambda(1+\lambda^2/16)\simeq\lambda_k$, with $\lambda_k$ the RWA value, so that the corrected amplitude is slightly smaller than the RWA one. Moreover, the shortening of the operation time by the factor $(1+\lambda^2/16)^{-1}$ is marginal, so the main benefit of the corrected parameters is the gain in fidelity, which in turn allows the use of larger amplitudes and, hence, faster gates.
 %-------------------------------------
 \section{Estimation of RWA inaccuracy}
The loss of accuracy in the gate representations, when the RWA-prescribed parameters $\Delta=0$, $\omega_0$, $\tau_g$ and $A/2$ [eqs. (\ref{eq:gate-times})] are used in numerical calculations, can be estimated as the infidelity resulting from replacing them into $F$, eqs. (\ref{eq:fidelity}), since the corrected parameters [eqs. (\ref{eq:gate-times-corrected})] are optimized for propagator $U(t)$. We use the following approximations: 

(i) \textit{Phase factors} $\sin(\omega t_g/2)$ are evaluated as $\sin(\omega_0\tau_g/2)$. Nevertheless, both $\sin(\omega t_g/2)=\pm 1$ [eqs. (\ref{eq:gate-times-corrected})], and $\sin(\omega_0 \tau_g/2)=\pm 1$ [eqs. (\ref{eq:gate-times})]; hence, $F$ only depends on $\Omega, \tau_g$ and $\theta$. 

(ii) \textit{Rabi frequency and operation time:} for the RWA parameters ($\Delta=0$), eq. (\ref{Rabi-correction}) gives $\Omega=(A/2)(1-\lambda^2/32)$, so that $\Omega t_g=x+\epsilon$ with $x=A\tau_g/2$ and $\epsilon=-A\tau_g\lambda^2/64\ll1$. Here $x=n\pi,\,(2n+1)\pi/2,\,(4n+1)\pi/4$ for $Z$, $X$ and $H$ (RWA conditions, eqs. (\ref{eq:gate-times})), i.e., $\cos x=\pm 1, 0, \pm1/\sqrt{2}$. Then, using $\cos (x+\epsilon)=\cos x \cos \epsilon-\sin x \sin \epsilon$ through lowest order, results 
\begin{equation}
    \label{eq:cos2}
    \cos^2(x+\epsilon) = \left\{ 
    \begin{array}{ll}
    \cos^2\epsilon \approx (1-\epsilon^2), &  Z, \\
        \sin^2\epsilon \approx \epsilon^2, &  X, \\
        (\cos\epsilon-\sin\epsilon)^2/2 \approx (1-2\epsilon)/2, &  H.
    \end{array}
    \right.
\end{equation}
%---------------------------
(iii) \textit{Quantization axis tilt:} the factors $\cos\theta=h_z/\Omega$ are evaluated with eqs. (\ref{eqs:hx-hy-hz}) at $\Delta=0$: $\cos\theta\approx(3A\lambda/8)/(A/2)=3\lambda/4$; hence $\sin\theta\approx 1-9\lambda^2/32$. \\

Finally, inserting approximations (i)--(iii) into eqs. (\ref{eq:fidelity}), the infidelities $I=1-F$ become, to lowest order in $\lambda$, 
\begin{subequations}
\label{power-law-error}
\begin{align}
I_Z &= \frac{2}{3}\epsilon^2=\frac{2}{3}\left(\frac{n\pi}{32}\right)^2 \lambda^4,  \\
I_X &\approx \frac{2}{3}\cos^2\theta=\frac{3}{8}\lambda^2, \\
I_H &\approx \frac{1}{3}\cos^2\theta=\frac{3}{16}\lambda^2,
\end{align}
\end{subequations}
where $n$ is the integer of eqs. (\ref{eq:gate-times}) ($|\epsilon|=n\pi\lambda^2/32$ for $Z$ because $A\tau_Z=2n\pi$). The $\epsilon$-dependent contributions to $I_X$ and $I_H$ are of order $\lambda^4$ and, henceforth, are neglected: the leading error of $X$ and $H$ comes from the tilt $\theta$, which does not depend on $n$, whereas $\epsilon$ grows with $n$ through $\tau_g$.

Equations (\ref{eq:fidelity}) show two distinct sources of error for gates designed with RWA parameters. The first is a tilt of the effective field away from the $-y$ axis ($\cos\theta\neq0$), caused by the mismatch between resonant driving ($\Delta=0$) and the optimal detuning $\Delta_{\rm BS}$, with $\cos\theta={\cal O}(\lambda)$. The second is an error $\epsilon$ in the rotation angle $\Omega t_g$, of order $\lambda^2$. Since $\sin(\omega t_g/2)=\pm1$, $F_Z$ is insensitive to the tilt and only the angle error contributes, so that $I_Z\propto\epsilon^2\propto\lambda^4$. In contrast, $F_X$ and $F_H$ are also sensitive to the tilt, which gives an infidelity of order $\cos^2\theta\sim\lambda^2$ that dominates at small $\lambda$. Therefore, the amplitude-dependent shift $\Delta_{\rm BS}$ explains the power-law deterioration of the fidelity when using resonant RWA-defined gate time $t_{\rm RWA}$ for arbitrary $A$, like the one shown in Fig. \ref{fig:Num-infidelity-lambda}. The dependence of $I_Z$ on $n$ reflects the longer operation time $\tau_Z\propto n$, so that the lowest dips correspond to $n=1$. 
%======================================
\section{Numerical calculations with corrected parameters}
We solved numerically the Schr\"odinger equation for $H_{\rm BS}(t)$ with the corrected parameters, eqs. (\ref{eq:gate-times-corrected}), to obtain the evolution operator $U$, and used it to represent the $Z$ and $H$ gates. 
%==================================
\begin{figure*}[h]
    \centering
    \begin{minipage}[c]{0.40\textwidth}
      \centering
      {\rm (a)}\\[2pt]
      \includegraphics[width=\linewidth]{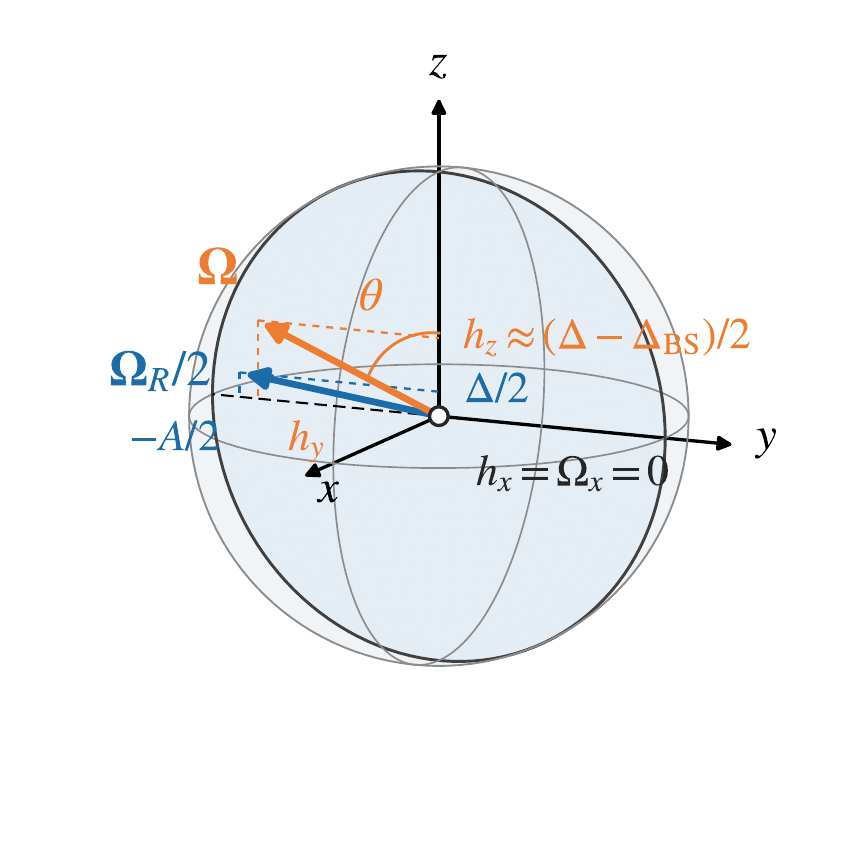}
    \end{minipage}\hspace{0.01\textwidth} \hspace{-2cm}
    \begin{minipage}[c]{0.565\textwidth}
      \centering
  $  \begin{array}{cc}
  {\rm (b)} & {\rm (c)} \\ 
    \includegraphics[width=0.4\textwidth]{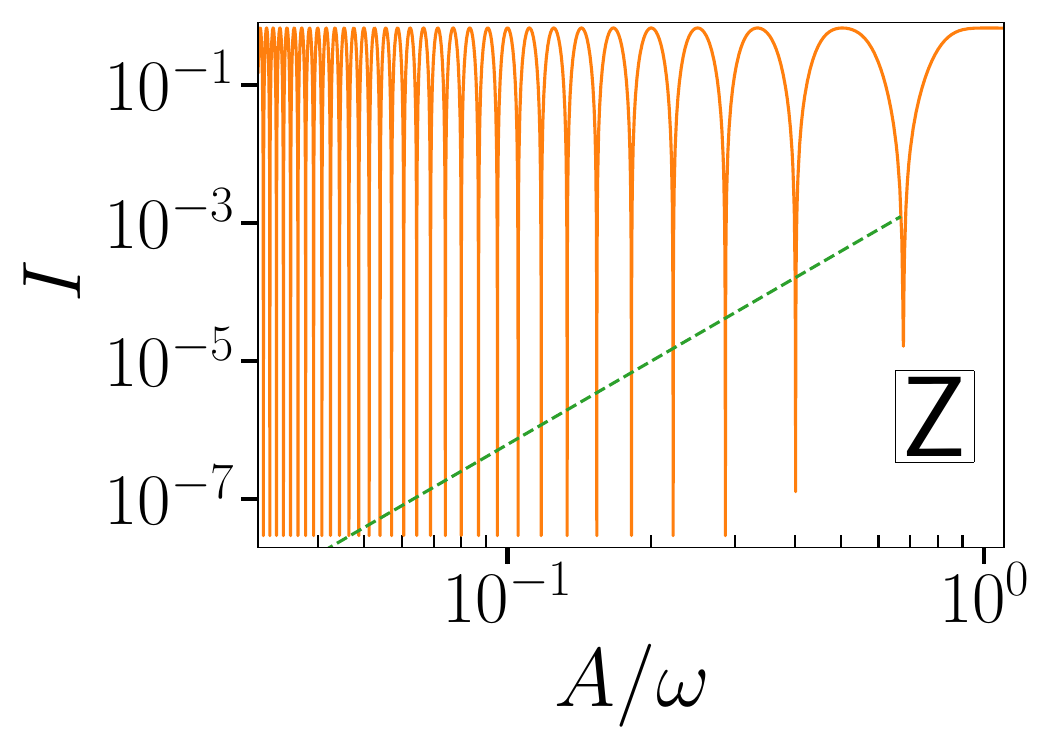} 
     & \hspace{-0.3cm} \includegraphics[width=0.4\textwidth]{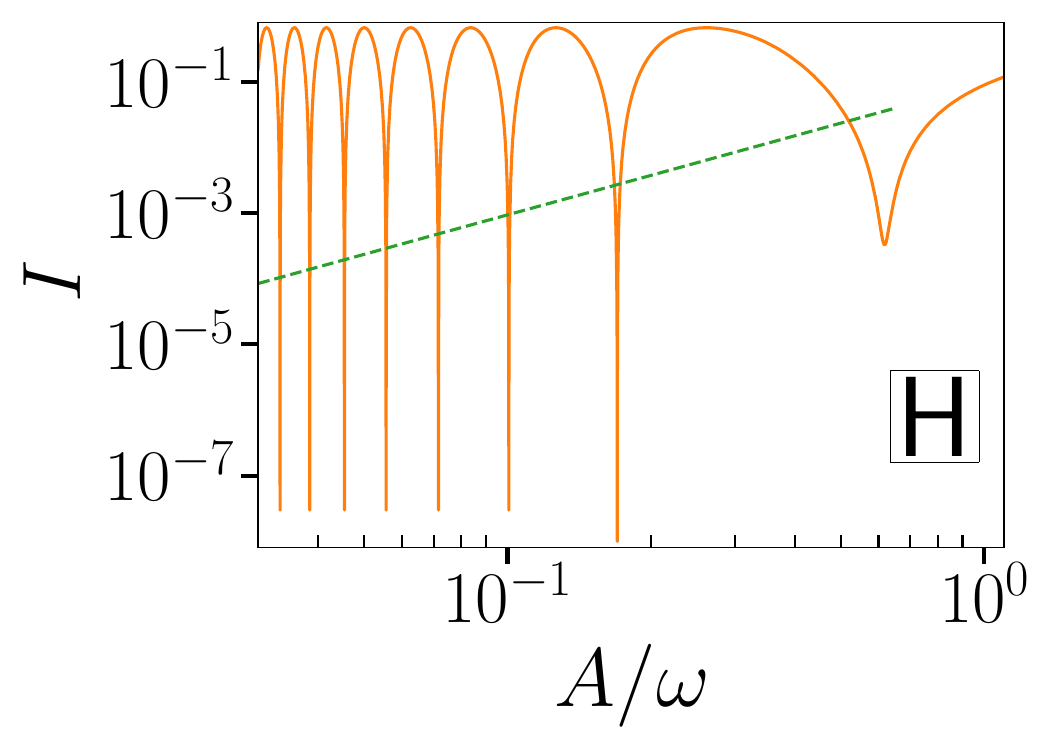} \\
     {\rm (d)} & {\rm (e)} \\
    \hspace{-0.3cm} \includegraphics[width=0.4\textwidth]{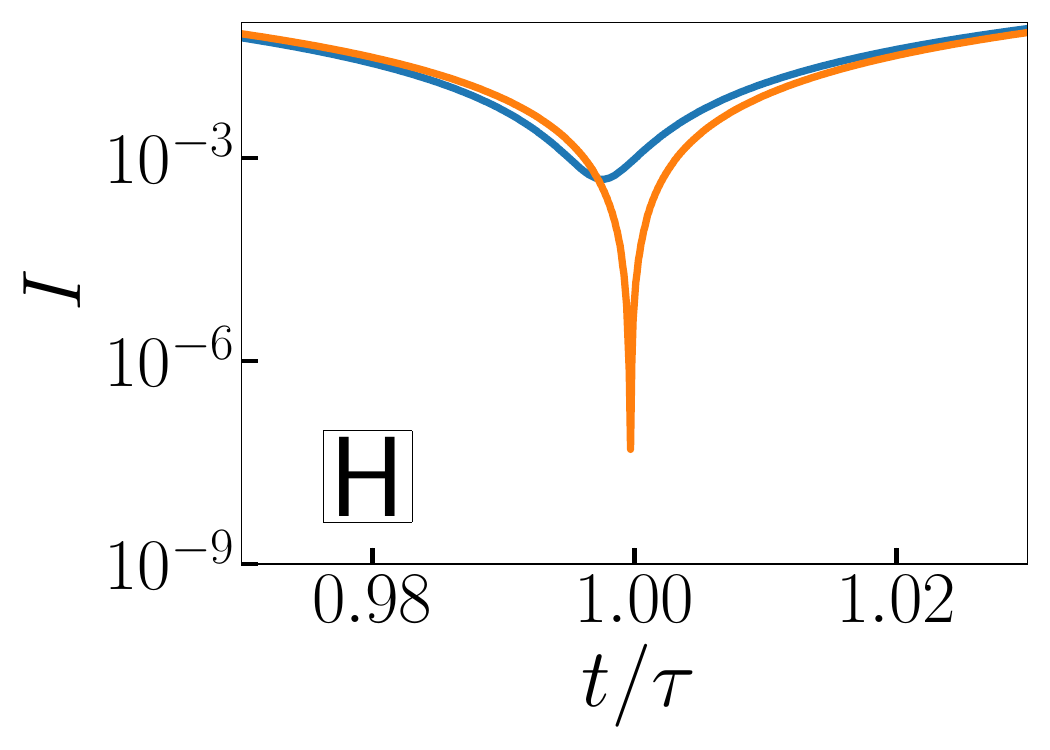} 
    & \hspace{-0.3cm}  \includegraphics[width=0.4\textwidth]{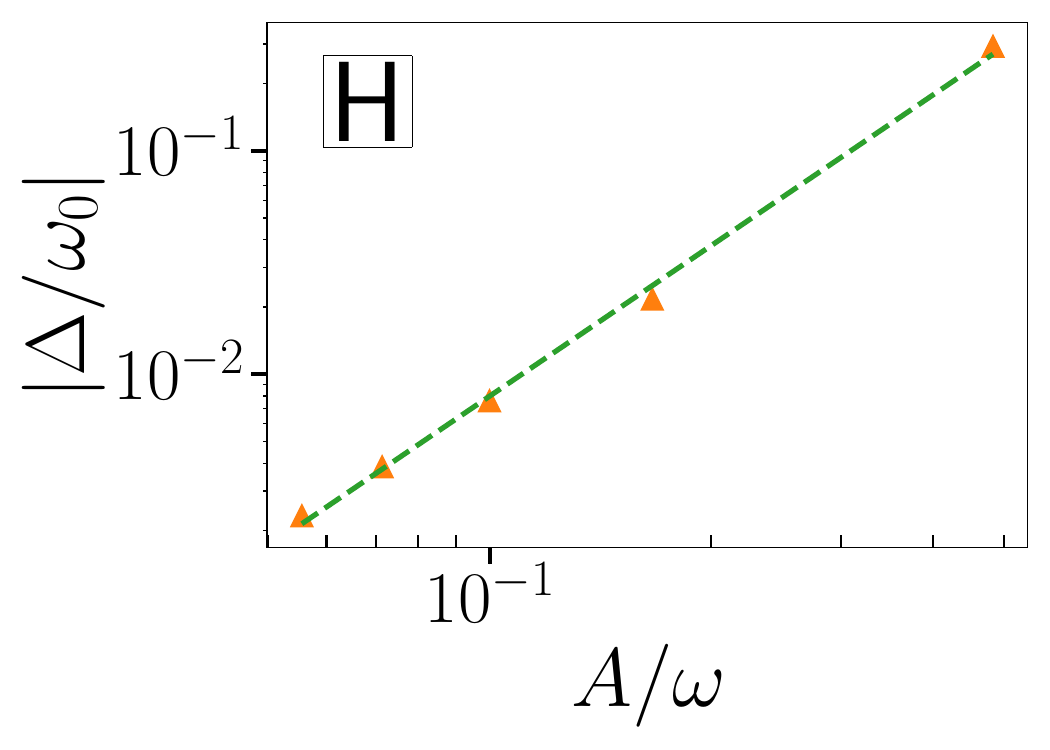} 
    \end{array} $
    \end{minipage}
    \caption{{\rm (a)} Equivalent magnetic fields acting on a spin 1/2, drawn on the Bloch sphere. Both $\mathbf{\Omega}$ and $\mathbf{\Omega}_R$ lie in the $(y,z)$ plane (shaded). Blue: field associated to the RWA, $\mathbf{\Omega}_R/2=(0,-A/2,\Delta/2)$. Orange: field of the Magnus-Taylor effective Hamiltonian, $\mathbf{\Omega}=(0,h_y,h_z)$, eqs. (\ref{eqs:hx-hy-hz}); its component $h_z\approx(\Delta-\Delta_{\rm BS})/2$ tilts it by an angle $\theta$ with respect to the $z$ axis, $\cos\theta=h_z/\Omega$. Dashed lines: components along $y$ and $z$; the drawing corresponds to $\Delta>0$ and $\Delta_{\rm BS}<0$, with exaggerated angles. Infidelity of the {\rm (b)} $Z$ and {\rm (c)} $H$ gates as a function of $A/\omega$, driving detuned by $\Delta_{\rm BS}$ with the corrected operation time, eqs. (\ref{eq:gate-times-corrected}). Dashed lines: power-law followed by the dips of the RWA gates drawn for comparison. {\rm (d)} Infidelity of the $H$ gate at $A/\omega=1/14$ as a function of the operation time in units of the RWA time $\tau_H$, same color code as {\rm (a)}. {\rm (e)} Numerically optimal relative detuning $|\Delta|/\omega_0$ for the $H$ gate as a function of $\lambda=A/\omega$ (symbols) and its fit (dashed line).}
    \label{fig:time-frequency-shift}
\end{figure*}
%====================
Figs. \ref{fig:time-frequency-shift}b and \ref{fig:time-frequency-shift}c show the calculated infidelity for the $Z$ and $H$ gates, obtained with the corrected parameters, eqs. (\ref{eq:gate-times-corrected}), as a function of $A/\omega$, as compared to those obtained with the RWA parameters at resonance represented by the same green dashed line as in Fig. \ref{fig:Num-infidelity-lambda}. Each dip corresponds to one of the amplitudes allowed by eqs. (\ref{eq:gate-times}). The infidelity of the corrected gates reach much lower values over the range of amplitudes shown.

Fig. \ref{fig:time-frequency-shift}d shows the $H$ gate at the particular dip $A/\omega=1/14$ as a function of the operation time, in units of the RWA time $\tau_H$. At resonant driving the numerically calculated infidelity has a broad minimum of order $10^{-3}$ (in agreement with the analytical second-order expression (\ref{power-law-error}c)), whereas introducing the detuning $\Delta_{\rm BS}$, eq. (\ref{BS-shift}), and the corrected operation time $t_g$, eq. (\ref{eq:gate-times-corrected}), produces a sharp minimum at least three orders of magnitude lower.

Fig. \ref{fig:time-frequency-shift}e shows the relative frequency shift that optimizes the $H$ gate, obtained numerically as a function of $\lambda=A/\omega$. The data fit to $\delta= -0.7 (A/\omega)^{2.0}\approx 3\lambda^2/4$, i.e. close to its analytical dependence, which supports the interpretation of $\Delta_{\rm BS}$ as the optimal gate detuning.
%========================================
\section{Conclusions}
We have studied the representation of the $Z$, $X$ and Hadamard gates by the unitary evolution operator of the second-order effective Magnus-Taylor Hamiltonian of a harmonically driven two-level system. 
Compared with the lowest-order RWA protocol, the dominant correction can be absorbed in two parameters, an amplitude-dependent shift of the driving frequency and a rescaling of the operation time, so that the RWA gate conditions remain valid with corrected parameters. 
The loss of fidelity of RWA-designed gates follows from the mismatch with these parameters and grows as $\lambda^4$ for $Z$ and as $\lambda^2$ for $H$. 
Notably, in spite of starting from a perturbative scheme, the numerical calculations agree with the analytical expressions in the range $A/\omega \lesssim 1$ explored here, and the numerically optimal detuning follows $\delta\approx-0.7\lambda^{2.0}$, close to eq. (\ref{BS-shift}). 
Nevertheless, it should be kept in mind that the detuning $\Delta_{\rm BS}$ is the optimal gate detuning for a drive with fixed initial phase and gates evaluated at stroboscopic times, not the shift of the spectral resonance and the residual error at large $\lambda$ is set by terms of order $A\lambda^3$ and by the truncation at ${\cal O}(1/\omega^3)$. 

Significant improvement of the fidelity is achieved by setting the operation time at $t_g\approx t_{\rm RWA}(1-A^2/16\omega_0^2)$ and operating off-resonant by detuning the driving frequency by the $\Delta_{\rm BS}$ shift, $\omega\approx\omega_0(1+3A^2/4\omega_0^2)$. Robustness against calibration errors and noise remains to be analyzed.
%%%%%%%%%%%%%%%%%%%
\section{Acknowledgments} 
Authors acknowledge financial support from CONICET (Argentina), and grants PI-20T001 and PI-23F005 (Universidad Nacional del Nordeste, Argentina).
%%%%%%%%%%%%%%%%%%%%%%%%%%%%%%%%%%%%%%

\acknowledgments
%================================0
\section{Appendix: Magnus-Taylor Hamiltonian up to ${\cal O}(1/\omega^2)$ \label{Magnus-Taylor Hamiltonian}}
For easy reference, we write here the explicit form of the effective Magnus-Taylor Hamiltonian $\widetilde{H}=H_0+H_1+H_2$ as given in Ref. \cite{DiVincenzo20}, with a phase offset $\phi=-\pi/2$ chosen for our problem
\begin{eqnarray*}
    H_0 &=& \frac{\Delta}{2}\sigma_z - \frac{A}{2}\sigma_y, \\
    %--------------------------------------
    H_1 &=& \frac{\Delta A}{4\omega}\left(\sin\beta_0 \sigma_x+\cos\beta_0 \sigma_y\right) + \frac{A^2}{8\omega}\left(1+2\cos\beta_0\right)\sigma_z, \\
    %---------------------------------------
    H_2 &=& \frac{A ^3}{32 \omega ^2} \left[\left(\sin 2 \beta _0-2 \sin \beta_0\right) \sigma _x \right. \\
   && + \left. \left(2 \cos \beta _0+\cos 2 \beta _0+2\right) \sigma _y\right] \\
   && -\frac{\Delta ^2 A}{8 \omega ^2}  \left(\sin \beta _0 \sigma _x  + \cos \beta _0 \sigma_y\right) -\frac{\Delta  A ^2}{8 \omega ^2} \left(1+\cos \beta _0\right) 
   \sigma _z.
\end{eqnarray*}
%================================
Then, separating the various terms depending on $\sigma_i$ ($i=x, y, z$), $\widetilde{H}$ is written in the main text as  $\widetilde{H} = h_x\sigma_x +h_y\sigma_y +h_z\sigma_z$, where 
%============================
\begin{eqnarray}
    h_x &=&  \frac{A}{2}\left(\lambda^2 c_x +\frac{\delta}{2}\sin\beta_0 -\frac{\delta^2}{4} \sin \beta _0 \right) \label{hx-general}\\ 
    %==================================
    h_y  &=& - \frac{A}{2} \left( 1  - \lambda^2 c_y- \frac{\delta}{2}\cos\beta_0  +\frac{\delta^2 }{4}  \cos \beta _0  \right)   \label{hy-general}   \\
    %-----------------------------------------------
   h_z &=& \frac{A}{2}\lambda c_z +\frac{\Delta}{2}\left[  1   -\frac{\lambda^2}{4} \left(1+\cos \beta _0\right)\right], \label{hz-general}
\end{eqnarray}
%---------------------------
where we used the shorthands
\begin{eqnarray}
    c_x &=& \frac{1}{16} \left(\sin 2 \beta _0-2 \sin \beta_0\right), \\
    c_y &=&  \frac{1}{16} \left(2 \cos \beta _0+\cos 2 \beta _0+2\right). \\
    c_z &=& \frac{1}{4}\left(1+2\cos\beta_0\right)
\end{eqnarray}
so that $h_i$ reduce for the resonant case ($\Delta=0$) to $h_x=c_x\lambda^2 A/2$, $h_y=-A/2(1-c_y \lambda^2)$ and $h_z=c_z\lambda A/2$.

Furthermore, expanding $\Omega$ [eq. (\ref{Omega-def})] up to ${\cal O}(1/\omega^2)$, and noticing that from the definition of the parameter $\lambda$, $\Delta/\omega\sim{\cal O}(\lambda^2)$ and  $\Delta^2/A^2\sim{\cal O}(\lambda^2)$,
\begin{eqnarray}
    \Omega &=& \frac{\Omega_R}{2} +\frac{\Delta  A ^2 }{8\Omega_R\omega} -\frac{A ^2 \left(4 \Delta ^4+6 \Delta ^2 A ^2+A ^4\right)}{64 \Omega_R^3\omega^2} \nonumber \\
    &=& \frac{\Omega_R}{2}\left(1-\frac{\lambda^2}{32}+\frac{A^2\Delta}{4\omega\Omega_R^2} -\frac{3A^2\Delta^2}{32\omega^2\Omega_R^2}\right) \nonumber \\
    &\approx& \frac{A}{2}\left(1+\frac{\Delta^2}{2A^2} \right) \left(1-\frac{\lambda^2}{32}+\frac{A^2\Delta}{4\omega\Omega_R^2} -\frac{3A^2\Delta^2}{32\omega^2\Omega_R^2}\right)
    \label{Omega-2nd-order}
\end{eqnarray}
where $\Omega_R=(\Delta^2+A^2)^{1/2}$ is twice the Rabi frequency, from which eq. (\ref{Rabi-correction}) of the main text follows.
%================================


\begin{thebibliography}{99}
\bibitem{note} In Ref. \cite{DiVincenzo20} the Hamiltonian is defined with a time-dependent amplitude $H_1(t)$ and a phase offset $\phi$.  Here we take $\phi=-\pi/2$ and a time-independent amplitude $H_1=A$ in order to compare to the quantum gates for harmonically-modulated tunneling qubit considered in \cite{Bendersky26}.
\end{thebibliography}
\end{document}